\documentclass[PRB,twocolumn,groupedaddress,nofootinbib]{revtex4-1}
\usepackage{graphicx}% Include figure files
\usepackage{dcolumn}% Align table columns on decimal point
\usepackage{bm}% bold math
\usepackage{tabularx}

\begin{document}

\title{Short- and long-range magnetic order in LaMnAsO}

\author{Michael A. McGuire}
\email{McGuireMA@ornl.gov}
\affiliation{Materials Science and Technology Division, Oak Ridge National Laboratory, Oak Ridge, Tennessee 37831 USA}
\author{V. Ovidiu Garlea}
\affiliation{Quantum Condensed Matter Division, Oak Ridge National Laboratory, Oak Ridge, Tennessee 37831 USA}

\begin{abstract}

The magnetic properties of the layered oxypnictide LaMnAsO have been revisited using neutron scattering and magnetization measurements. The present measurements identify the N\'{e}el temperature $T_N$\,=\,360(1)\,K. Below $T_N$ the critical exponent describing the magnetic order parameter is $\beta$\,=\,0.33$-$0.35, consistent with a three dimensional Heisenberg model. Above this temperature, diffuse magnetic scattering indicative of short-range magnetic order is observed, and this scattering persists up to $T_{SRO}$\,=\,650(10)\,K. The magnetic susceptibility shows a weak anomaly at $T_{SRO}$ and no anomaly at $T_N$. Analysis of the diffuse scattering data using a reverse Monte Carlo algorithm indicates that above $T_N$ nearly two- dimensional, short-range magnetic order is present with a correlation length of 9.3(3)\,{\AA} within the Mn layers at 400\,K. The inelastic scattering data reveal a spin-gap of 3.5\,meV in the long-range ordered state, and strong, low-energy (quasi-elastic) magnetic excitations emerging in the short-range ordered state. Comparison with other related compounds correlates the distortion of the Mn coordination tetrahedra to the sign of the magnetic exchange along the layer-stacking direction, and suggests that short-range order above $T_N$ is a common feature in the magnetic behavior of layered Mn-based pnictides and oxypnictides.

\end{abstract}

\maketitle

\section{Introduction}
The study of layered iron pnictide and chalcogenide superconductors has led to the discovery of other interesting properties in isostructural analogues. The structure types adopted by these materials are amenable to a variety of chemical substitutions, which allow access to a wide range of physical behaviors, including multiple types of superconducting, magnetic, and electronic states. This is especially true of the 1111-type materials with the ZrCuSiAs structure type \cite{Pottgen-2008} adopted by the prototype iron-based, high-temperature superconductor parent phase LaFeAsO \cite{Jeitschko-2000}.

Soon after the discovery of superconductivity in F-doped LaFeAsO \cite{Kamihara-2008}, substitution of La with other rare-earth ions (\textit{Ln}$^{3+}$) was seen to significantly increase the superconducting transition temperature \cite{Chen-2008, Ren-2008}, and to induce superconductivity in the case of partial substitution of Th$^{4+}$ \cite{Wang-2008}. Arsenic can be replaced by antimony in the case of \textit{Ln}MnSbO and \textit{Ln}ZnSbO \cite{Schellenberg-2008}, or by phosphorus, producing low temperature superconductors \textit{Ln}FePO with anisotropic gap structures \cite{Kamihara-2006, Fletcher-2009}. Oxygen can be fully replaced by fluorine to form compounds like \textit{A}FeAsF (\textit{A}$^{2+}$\,=\,Ca, Sr), whose properties are quite similar to those of \textit{Ln}FeAsO, including coupled structural and antiferromagnetic phase transitions and superconductivity when doped, due to the presence of isoelectronic FeAs layers \cite{Tegel-2008, Matsuishi-2008a, Matsuishi-2008b, Han-2008}. Doping other transition metals on the Fe site can induce superconductivity \cite{Sefat-2008}, while full replacement of Fe produces diverse and interesting phenomena. These include giant magnetoresistance and rare-earth magnetism in \textit{Ln}MnAsO \cite{Marcinkova-2010, Emery-2010, Tsukamoto-2011, Wildman-2012}, itinerant ferromagnetism in \textit{Ln}CoAsO \cite{Yanagi-LaCoAsO, Ohta-LCoAsO, McGuire-2010, Marcinkova-NdCoAsO}, low temperature superconductivity and Kondo physics in \textit{Ln}NiAsO \cite{Watanabe-LaNiAsO, Li-LaNiAsO, Matsuishi-LnNiAsO, Luo-CeNiAsO}, and glassy magnetism in \textit{Ln}RuAsO \cite{McGuire-2014}.

Of particular interest is the series of Mn-based 1111 materials \textit{Ln}MnAsO, which are antiferromagnetic (AFM) insulators exhibiting giant magnetoresistance \cite{Emery-2010, Wildman-2012} and, when doped, metal-insulator transitions \cite{Sun-2012, Hanna-2013} and low temperature crystallographic distortions \cite{Wildman-2015}. Electronic properties of LaMnAsO thin films indicate p-type conduction and a band gap of 1.0$-$1.4\,eV \cite{Kayanuma-2009}, and calculations suggest that LaMnAsO$-$LaZnAsO alloys may provide a source of spin-polarized carriers for spintronic applications \cite{Li-2013}.

CeMnAsO, PrMnAsO$_{0.95}$F$_{0.05}$, and NdMnAsO have been shown via neutron diffraction to have antiferromagnetic ordering of Mn moments below $T_N$ = 347, 340 and 359\,K, respectively \cite{Zhang-2015, Wildman-2015, Marcinkova-2010}. Powder neutron diffraction measurements on LaMnAsO have shown AFM order at 300\,K, and the absence of magnetic Bragg reflections at 400\,K \cite{Emery-2011}. Based on magnetization measurements, values of 317$-$325\,K have been reported for the $T_N$ of LaMnAsO \cite{Emery-2010, Sun-2012}.

Like \textit{Ln}MnAsO, the analogous 122 material BaMn$_2$As$_2$ is also an antiferromagnetic Mott insulator \cite{An-2009, Singh-2009a, Singh-2009b, Johnston-2011}, and dilute magnetic semiconductor behavior has been observed in materials produced by simultaneously alloying K and Zn on the the Ba and Mn sites, respectively \cite{Zhao-2013}. Magnetization measurements \cite{Johnston-2011} and neutron diffraction \cite{Singh-2009b} have determined the N\'{e}el temperature ($T_N$) of BaMn$_2$As$_2$ is near 620\,K, significantly higher than $T_N$ for \textit{Ln}MnAsO. The magnetic anomaly at $T_N$ in BaMn$_2$As$_2$ is subtle, and the susceptibility does not obey a Curie Weiss law above this temperature. These observations suggest magnetic correlations exist above the long-range ordering temperature \cite{Johnston-2011}.

We recently synthesized LaMnAsO in our laboratory \cite{Idrobo-2015} and found no magnetic anomaly near the reported N\'{e}el temperature, indicating that the presence of ferromagnetic MnAs ($T_C$ = 318\,K \cite{Guillaud-1951, Suzuki-1982, Wada-2001}) as an secondary phase in some samples is likely responsible for the large magnetic anomaly usually observed near this temperature \cite{Sun-2012}. We performed high temperature magnetization measurements and observed a subtle anomaly near 650\,K, reminiscent of the feature observed at $T_N$ = 618\,K in Ref. \citenum{Johnston-2011} for BaMn$_2$As$_2$. However, as noted above, a previous neutron diffraction study showed that $T_N$ of LaMnAsO is below 400\,K \cite{Emery-2011}.

To resolve the uncertainty regarding the true N\'{e}el temperature of LaMnAsO and understand the nature of the magnetic anomaly that we observe at high temperature, we complemented our magnetization study with temperature dependent powder neutron diffraction and inelastic scattering measurements. Here we report the findings of these investigations, which include long-range magnetic order below $T_N$ = 360(1)\,K and short-range magnetic order above $T_N$ persisting up to about 650\,K. We also observe intense quasi-elastic scattering in the short range ordered state, and strongly dispersing magnetic excitations in both the short- and long-range ordered states. A spin gap of about 3.5 meV is measured at room temperature. These findings provide a better understanding of the magnetic nature of LaMnAsO that appears to extend to other compounds with similar [MnAs]$^{1-}$ layers.

\section{Experimental Details}
LaMnAsO powder samples were made by first reacting freshly filed La powder with As pieces to form LaAs. This was combined in a 1:1 molar ration with MnO and pressed into pellets. The pellets were heated at 1000\,$^{\circ}$C for 48\,h, ground and re-pelletized, and heated again at 1000\,$^{\circ}$C for 48\,h. All reactions were done inside evacuated and flame sealed silica-glass ampoules.

Powder x-ray diffraction (PXRD) data were collected using a PANalytical X'Pert Pro MPD diffractometer with monochromatic Cu K$_{\alpha1}$ radiation. An Oxford PheniX cryostat was used for PXRD below room temperature. Rietveld refinement of room temperature PXRD data indicated the sample used in this study was 94\% pure with 3\% each of MnO and La$_2$O$_3$ by weight. Energy dispersive spectroscopic analysis carried out using a Bruker Quantax 70 x-ray spectrometer in a Hitachi TM3000 tabletop scanning electron microscope gave a composition of La$_{26(1)}$Mn$_{26(1)}$As$_{25(1)}$O$_{23(1)}$ determined by averaging multiple measurements on different crystallites in the sample, consistent with the expected 1111 stoichiometry. Magnetization measurement were performed with a Quantum Design Magnetic Property Measurement System, with the sample sealed inside a silica tube with approximately 1/3 of an atmosphere of ultra-high purity argon gas.

Neutron diffraction data was collected from 5\,g of LaMnAsO powder at the High Flux Isotope Reactor at Oak Ridge National Laboratory (ORNL) using the Neutron Powder Diffractometer (beamline HB-2A). The sample was held in a cylindrical vanadium container, loaded in a closed cycle refrigerator capable of reaching temperatures from 6 K to 800 K. Measurements were performed using the wavelengths $\lambda$ = 1.541 and 2.414 $\AA$ provided by a vertically focusing Ge monochromator. For data collection, a detector array consisting of 44 $^3$He tubes was scanned to cover the total 2$\theta$ range of 7$-$150$^{\circ}$ in steps of 0.05$^{\circ}$. Inelastic neutron scattering (INS) data was collected using the Hybrid Spectrometer (HYSPEC) at the Spallation Neutron Source at ORNL. HYSPEC is a direct geometry spectrometer that combines time-of-flight spectroscopy with focusing Bragg optics. The incident neutron beam is monochromated using a Fermi chopper and is then vertically focused by Bragg scattering from a monochromator onto the sample position.  HYSPEC employs $^3$He linear position sensitive tube detectors that are assembled into 20 sets of 8-packs that cover an angular range of 60$^{\circ}$ in the horizontal scattering plane and a vertical acceptance of 15$^{\circ}$ \cite{Winn-2015}. Data on LaMnAsO was collected using polarized neutrons with incident energies of $E_i$ = 15\,meV selected by a Heusler monochromator. A Mezei flipper was used to reverse the polarization direction of the incident beam, and the polarization analysis was done using a wide-angle supermirror bender. The polarization vector was oriented within the horizontal scattering plane by using a guide field generated from a set of orthogonal XYZ coils situated around the sample position. For the INS study the powder was loaded in aluminium cans and placed in a vacuum furnace.

Rietveld refinement of x-ray and neutron diffraction data was performed using Fullprof \cite{Fullprof}. Diffuse magnetic scattering was modeled using the reverse Monte Carlo algorithm employed by the program SPINVERT \cite{SPINVERT}.

\section{Results and Discussion}

\subsection{Crystal Structure and Thermal Expansion}

\begin{figure}
\begin{center}
\includegraphics[width=3.00in]{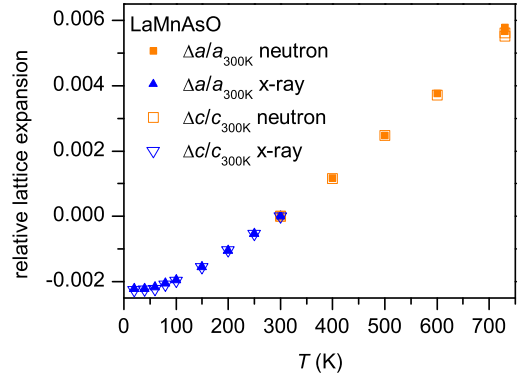}
\caption{\label{fig:latt}
Temperature dependence of the lattice parameters of LaMnAsO. Neutron and x-ray diffraction results are shown for $T \geq 300 K$ and $T \leq 300 K$, respectively. Each data set was separately normalized to 300\,K values. See also Table \ref{tab:npd}.
}
\end{center}
\end{figure}

Crystallographic details determined at 20, 300, and 730\,K from x-ray and neutron diffraction are collected in Table \ref{tab:npd}. The unit cell parameters and atomic positions agree well with previous reports \cite{Nientiedt-1997, Emery-2010}. Upon cooling from 300 to 20\,K, \textit{a} and \textit{c} both contract by 0.22\%, indicating remarkably isotropic thermal expansion in this tetragonal compound. Figure \ref{fig:latt} confirms this, showing the lattice constants, normalized to 300\,K values, determined from 20 to 730\,K using x-ray and neutron diffraction.

Calculation of the linear coefficient of thermal expansion using the data between 200 and 300\,K gives 1.05$\times 10^{-5}$\,K$^{-1}$ along \textit{a} and 1.03$\times 10^{-5}$\,K$^{-1}$ along \textit{c}. Nearly isotropic thermal expansion near room temperature is also seen in NdMnAsO \cite{Emery-2011}, where \textit{c} changes with temperature about 1.2 times as fast as \textit{a}, and PrMnAsO$_{0.95}$F$_{0.05}$ \cite{Wildman-2015}, where this ratio is 1.3. This can be compared to isostructural LaFeAsO \cite{McGuire-2008} and NdCoAsO \cite{McGuire-2010}, in which the ratios of linear thermal expansion along the different crystallographic directions \textit{c} and \textit{a} are 2.0 and 5.4, respectively, indicting that this anisotropy likely arises from the transition metal 3\textit{d}-orbitals.

Table \ref{tab:compare} lists the Mn$-$Mn interatomic distances determined near room temperature in LaMnAsO along with several related 1111 and 122 compounds that also contain layers of Mn in edge sharing tetrahedral coordination. Also shown is the $Pn-$Mn$-Pn$ bond angle $\alpha$ formed by to two neighboring pnictogen (\textit{Pn}) atoms located above a Mn layer. The angle $\beta$, formed by two \textit{Pn} on opposite sides of the Mn layer, is related to $\alpha$ by cos($\beta$)\,=\,-$\frac{1}{2}$[1+cos($\alpha$)]. For ideal tetrahedral geometry, both angles are 109.5 degrees. Long range antiferromagnetic ordering temperatures ($T_N$), magnetic moments on Mn determined at low temperature ($\mu_{Mn}$), and the magnetic structure defined by how the layers stack along the $c$ direction are also listed. Correlations between crystal structure and magnetic behavior observed in these data will be discussed below.

\begin{table}
\begin{center}
\caption{\label{tab:npd} Crystallographic parameters for LaMnAsO determined by Rietveld refinement of powder x-ray and neutron diffraction data using space group $P4/nmm$ with La at ($\frac{1}{4}$, $\frac{1}{4}$, \textit{z}$_{La}$), Mn at ($\frac{3}{4}$, $\frac{1}{4}$, 0), As at ($\frac{1}{4}$, $\frac{1}{4}$, \textit{z}$_{As}$), and O at ($\frac{3}{4}$, $\frac{1}{4}$, $\frac{1}{2}$).
}
\begin{tabularx}{0.9\columnwidth}{l|cccc}
\toprule												
$T$ (K)                   &  20	& 300      &	300	&         	730	\\
radiation                 &  x-ray    & x-ray      & neutron       & neutron \\
$\lambda$ ({\AA})         &  1.540598    & 1.540598  & 1.541        & 1.541 \\
\hline
\textit{a} ({\AA})	      &  4.1109(1)    & 4.1200(1)     &  4.1235(2)	&	4.1473(1)	\\
\textit{c} ({\AA})	      &  9.0259(3)    & 9.0462(2)	&  9.0532(5)	&	9.1040(3)	\\
$z_{La}$	              &  0.6328(3)    & 0.6323(3)	&  0.6325(4)	&	0.6319(3)	\\
$z_{As}$	              &  0.1680(5)   &  0.1684(4)	 &  0.1685(6)	&	0.1687(3)	\\
$B_{La}^{iso}$ ({\AA}$^2$)&	--  &   --	 &  0.45(8)	      &	0.79(6)	\\
$B_{Mn}^{iso}$ ({\AA}$^2$)&	--    &  --   &  0.58(14)	    &	1.34(12)	\\
$B_{As}^{iso}$ ({\AA}$^2$)& --    &  --  &  0.65(10)	      &	1.29(9)	\\
$B_{O}^{iso}$ ({\AA}$^2$)&	--   &      --	 &  0.62(10)	    &	1.01(9)	\\
\hline
$\chi^2$	              &  1.89   &  1.65         &	1.86	&	1.08	\\
\toprule													
\end{tabularx}
\end{center}
\end{table}
\begin{table*}
\begin{center}
\caption{\label{tab:compare} Comparison of Mn$-$Mn distances near room temperature and the angle ($\alpha$) formed by two coplanar pnictogen (\textit{Pn}) atoms and their shared Mn atom, long range magnetic ordering temperatures, ordered magnetic moments at low temperature, and stacking pattern of checkerboard AFM layers in some manganese pnictides containing layers of Mn in edge-sharing tetrahedral coordination.
}
\begin{tabularx}{1.55\columnwidth}{lccccccc}
\toprule													
compound	&	$d_{Mn-Mn}^{intralayer}$	&	$d_{Mn-Mn}^{interlayer}$	&	$\alpha$($Pn-$Mn$-Pn$) 	&	$T_N$	&	 $m_{Mn}$	&	stacking	&	Refs.	\\
	&	{\AA}	&	{\AA}	&	deg.	&	K	&	$\mu_B$	&		&		\\
\hline															
LaMnAsO	&	2.913	&	9.046	&	107.0	&	360(1)	&	3.34(2)	&	FM	&	this work, \citenum{Emery-2011}	\\
CeMnAsO	&	2.889	&	8.956	&	106.7	&	347(1)	&	3.32(4)	&	FM	&	\citenum{Nientiedt-1997, Zhang-2015}	 \\
PrMnAsO$_{0.95}$F$_{0.05}$	&	2.877	&	8.921	&	105.6	&	340	&	3.7	&	FM	&	\citenum{Wildman-2015}	\\
NdMnAsO	&	2.864	&	8.905	&	105.0	&	359	&	3.54(4)	&	FM	&	\citenum{Emery-2011}	\\
NdMnAsO$_{0.95}$F$_{0.05}$	&	2.863	&	8.897	&	105.2	&	356(2)	&	3.83(2)	&	FM	&	\citenum{Wildman-2012}	 \\
LaMnPO	&	2.869	&	8.843	&	111.2	&	$>$375	&	3.28(5)	&	FM	&	\citenum{Yanagi-2009}	\\
PrMnSbO	&	2.961	&	9.472	&	--	&	230	&	3.69(3)	&	FM	&	\citenum{Kimber-2010, Schellenberg-2008}	\\
BaMnAsF	&	3.022	&	9.588	&	110.3	&	338(1)	&	3.65(5)	&	AFM	&	\citenum{Saparov-2013}	\\
BaMnSbF	&	3.167	&	9.830	&	107.2	&	272(1)	&	3.66(3)	&	AFM	&	\citenum{Saparov-2013}	\\
BaMn$_2$As$_2$	&	2.948	&	6.734	&	108.7	&	625	&	3.88(4)	&	AFM	&	\citenum{Singh-2009b}	\\
BaMn$_2$P$_2$	&	2.854	&	6.526	&	111.8	&	--	&	4.2(1)	&	AFM	&	\citenum{Brock-1994}	\\
\toprule													
\end{tabularx}
\end{center}
\end{table*}

\subsection{Magnetization}

\begin{figure}
\begin{center}
\includegraphics[width=3.00in]{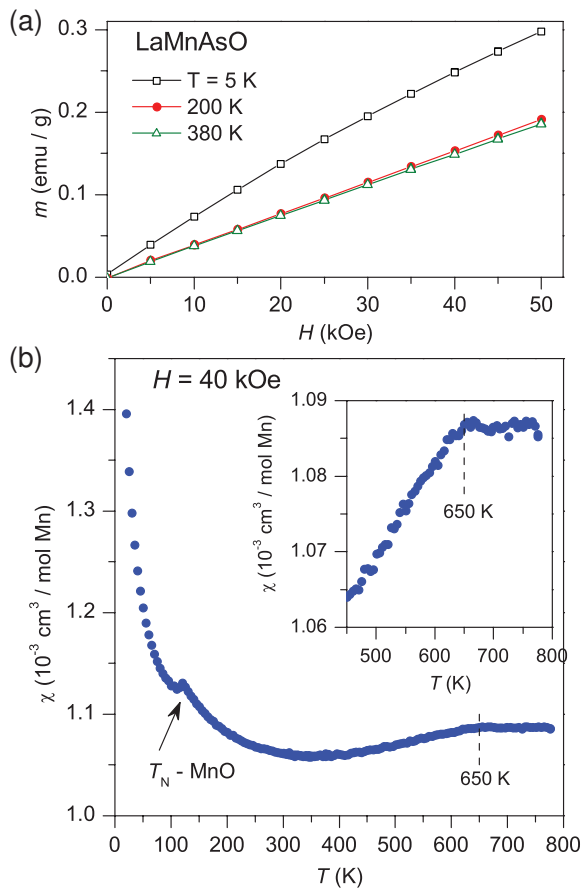}
\caption{\label{fig:mag}
Results of magnetization measurements on LaMnAsO. (a) Isothermal magnetization curves showing the absence of any significant ferromagnetic signal. (b) Temperature dependence of the magnetic susceptibility, with the anomaly at 650\,K indicated on the figure and the inset. Note the small cusp associated with the MnO impurity, and the absence of any magnetic anomaly near room temperature.
}
\end{center}
\end{figure}

Results from magnetization measurements are shown in Figure \ref{fig:mag}. Isothermal magnetization curves at 5, 200, and 380\,K (Fig. \ref{fig:mag}a) show no indication of significant ferromagnetic components. The \textit{H}\,=\,0 intercept extrapolated from the high field portion of the 200\,K \textit{M} vs \textit{H} data shows that the sample contains less than 0.001\% MnAs by weight \cite{Ido-1985}. The cusp in magnetic susceptibility ($\chi$) near 120\,K (Fig. \ref{fig:mag}b) arises from the small MnO impurity observed by diffraction, which orders antiferromagnetically at this temperature \cite{Tyler-1933}. The low temperature upturn upon cooling is expected to arise from a small concentrations of paramagnetic moments associated with defects or secondary phases. The Curie constant determined by fitting the low temperature tail corresponds to 0.03 spin 1/2 moments per formula unit.

No significant anomaly is observed in $\chi$ (Fig. \ref{fig:mag}b) between 300 and 400\,K, the temperature range in which $T_N$ is expected to lie \cite{Emery-2011}. A change in the temperature dependence of $\chi$ is observed at 650\,K. This similar to the anomaly observed at 618\,K in the closely related material BaMn$_2$As$_2$ \cite{Johnston-2011}, which was shown to coincide with the onset of long-range antiferromagnetic ordering of Mn moments \cite{Singh-2009b}. However, neutron diffraction results presented below will show that in LaMnAsO this high temperature feature arises from the development of short-range magnetic order, and long-range order occurs only upon cooling below 360\,K, with no associated anomaly in the magnetic susceptibility.

\subsection{Neutron scattering: magnetic order}

\begin{figure}
\begin{center}
\includegraphics[width=3.00in]{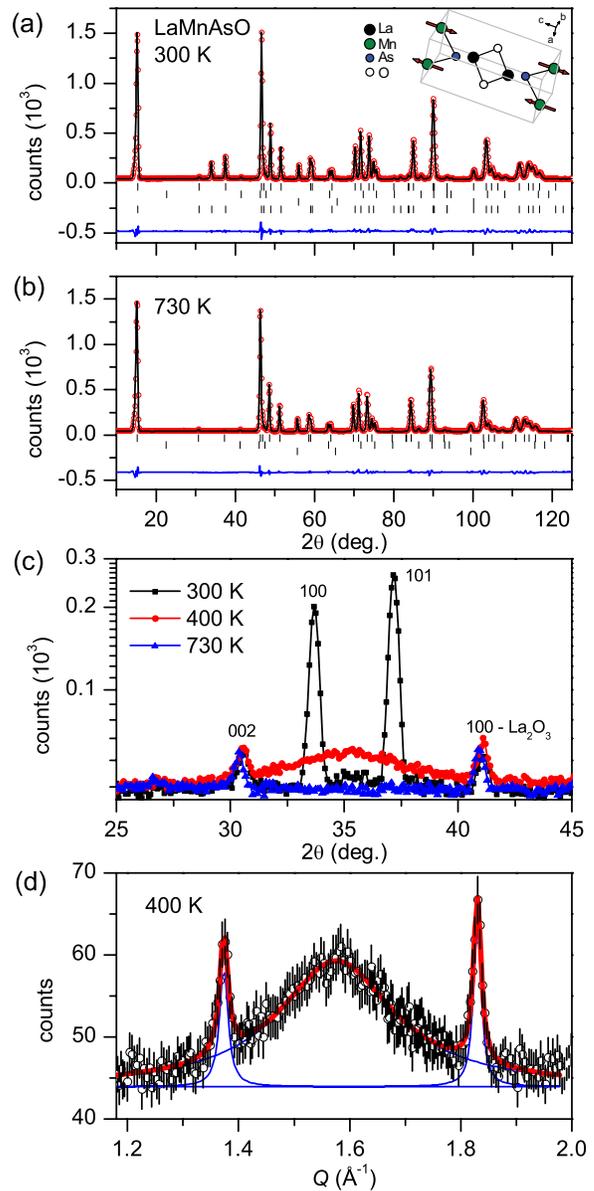}
\caption{\label{fig:npd}
Neutron powder diffraction data from LaMnAsO collected with a neutron wavelength of 2.414\,{\AA}. (a) Rietveld refinement of data collected at 300\,K. (b) Rietveld refinement of data collected at 730\,K. Upper set of ticks locate nuclear reflections, lowest set of ticks at 300\,K locate magnetic reflections, other sets indicate reflections from MnO and La$_2$O$_3$ impurities. (c) Comparison of diffraction patterns showing sharp peaks at 300\,K indicating long-range magnetic order, a broad peak at 400\,K indicating short-range order, and no magnetic scattering at 730\,K. (d) A fit to the short-range order peak using a Lorentzian function. The two sharp reflections that appear in the \textit{Q} range of interest are modeled in the same way.
}
\end{center}
\end{figure}

Neutron powder diffraction patterns and Rietveld refinement results at 300 and 730\,K are shown in Figure \ref{fig:npd}a and \ref{fig:npd}b, respectively. At 730\,K the data are fully described by the nuclear structures with no indication of magnetic scattering. Additional sharp Bragg reflections indicating long-range magnetic order (LRO) are apparent at 300\,K, and were modeled well using the magnetic structure first reported by Emery \textit{et al.} in Refs. \citenum{Emery-2010} and \citenum{Emery-2011}. The magnetic structure is shown in the inset of Figure \ref{fig:npd}a. Moments on Mn form a checkerboard pattern in the \textit{ab}-plane and these planes stack ferromagnetically along the \textit{c}-direction. Such a spin arrangement is described by the magnetic space group $P 4'/n'm'm$ (with Mn1 at (3/4,1/4,1/2 $|$ 0,0,$m_z$) and Mn2 at (1/4,3/4,1/2 $|$ 0,0,$-m_z$)). This same magnetic structure is adopted at room temperature by the other \textit{Ln}MnAsO compounds with \textit{Ln}\,=\,Ce, Pr, and Nd \cite{Marcinkova-2010, Zhang-2015, Wildman-2015}. The fit shown in Figure \ref{fig:npd} indicates a magnetic moment of 2.27(3)\,$\mu_B$ per Mn at 300\,K. As expected, this is slightly smaller than the value of 2.43(1)\,$\mu_B$ per Mn reported at 290\,K \cite{Emery-2010}.

Structural and magnetic properties of 1111 and 122 Mn pnictide compounds are collected in Table \ref{tab:compare}. In all of the compounds the ordered moment at low temperature saturates to values significantly smaller than the full moment of $gS = 5 \mu_B$, with values in the range of 3.2$-$4.2\,$\mu_B$. This moment reduction has been attributed to quantum fluctuations \cite{Johnston-2011} or Mn$-$As hybridization \cite{An-2009}. The largest magnetic moments are seen in 122 compounds. The long range ordering temperature is highest for BaMn$_2$As$_2$, which could be attributed to the shortest interlayer Mn$-$Mn distance; this is expected to be the direction of weakest magnetic exchange, and therefore should determine the onset of LRO. Within the 1111 type compounds there is no clear overall trend in $T_N$ with Mn$-$Mn distance, although the lowest $T_N$ values, for PrMnSbO and BaMnSbF, occur with two of the largest intralayer Mn$-$Mn separation. The most interesting correlation seen in the table is between this bond angle and the stacking sequence of the checkerboard-ordered Mn layers along the \textit{c} axis. With one exception (LaMnPO), all compounds with $\alpha \leq 107.0 ^{\circ}$ have layers stacked FM resulting in a C-type AFM structure, while those with $\alpha > 107.0 ^{\circ}$ stack AFM giving a G-type AFM structure. This shows that the magnetic exchange along the \textit{c}-axis is determined by the details of the crystal field splitting of the Mn 3\textit{d} orbitals. This suggests that increasing the angle $\alpha$ with applied or chemical pressure may switch the magnetic order from C-type to G-type in some 1111 Mn compounds.

Data collected at intermediate temperatures was used to determine the lattice parameters shown in Figure \ref{fig:latt}, and analyzed for indications of magnetic scattering. As expected based on Ref. \citenum{Emery-2011}, the sharp magnetic Bragg peaks were absent at 400\,K; however additional diffuse scattering centered near the magnetic 100 and 101 Bragg peaks was observed. This is shown Figure \ref{fig:npd}c, which compares diffraction patterns collected at 300, 400, and 730\,K. The diffuse scattering is not detectable at 730\,K, and nearly completely suppressed in the long-range ordered state at 300\,K. This is identified as a signature of short-range magnetic order (SRO) above the N\'{e}el temperature. An estimate of the correlation length ($\xi$) of the SRO was determined by fitting the \textit{Q} dependence of the intensity to a Lorentzian lineshape $L(Q) = L_0 + (2A/\pi)\cdot w/[4(Q-Q_0)^2 - w^2$], where $\xi = 2/w$. The fit is shown in Figure \ref{fig:npd}d, which gives $\xi$ = 8.0(3)\,{\AA}. This SRO is examined in more detail below.

\begin{figure}
\begin{center}
\includegraphics[width=3.00in]{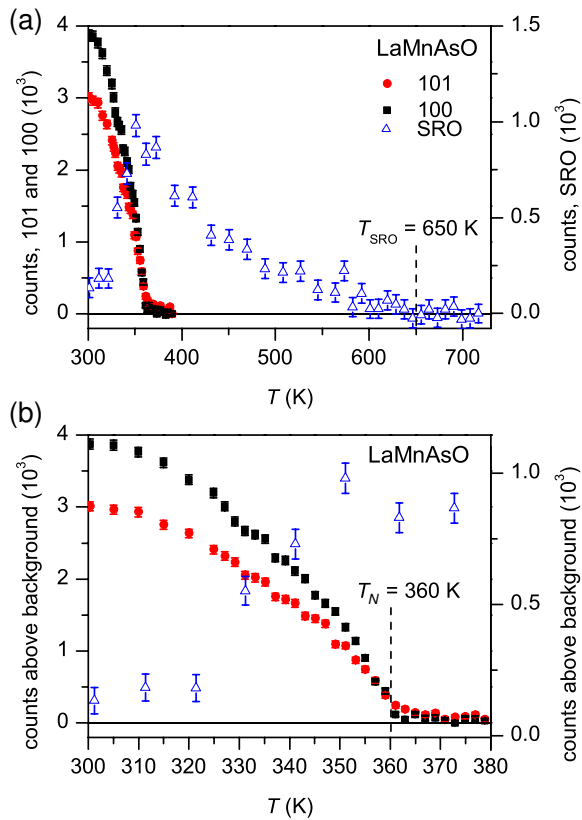}
\caption{\label{fig:op}
Temperature dependence of magnetic scattering from LaMnAsO. (a) Counts measured at the center of the 101, 100, and short-range order (SRO) peaks. (b) The behavior of the data shown in (a) near the N\'{e}el temperature.
}
\end{center}
\end{figure}

Order parameters were determined as a function of temperature for the LRO and SRO by counting with the detector at the center of the sharp and diffuse peaks in Figure \ref{fig:npd}d, respectively. The results are shown in Figure \ref{fig:op}. Upon heating from room temperature, the LRO is observed to vanish near 360\,K. Fits of the order parameter (defined as the square root of the measured intensity) above 330\,K to the functional form $(T_N-T)^\beta$ give $T_N$\,=\,361.4(5)\,K and $\beta$\,=\,0.35(2) using the 101 reflection and $T_N$\,=\,360.5(3)\,K and $\beta$\,=\,0.33(1) using the 100 reflection using either the 100 or 101 reflection. Thus, an N\'{e}el temperature of 360(1)\,K is determined. The values determined for the critical exponent $\beta$ are close to those expected for a three dimensional Heisenberg systems. For BaMn$_2$As$_2$ a similar $\beta$ value of 0.35(2) has been reported \cite{Singh-2009b}. Among Mn-based 1111 materials, reports include a significantly higher $\beta$\,=\,0.47 for CeMnAsO \cite{Zhang-2015} and a significantly lower $\beta$\,=\,0.27(1) for NdMnAsO \cite{Marcinkova-2010}. The fitting range used in Ref. \cite{Marcinkova-2010} for NdMnAsO extends to \textit{T}\,=\,0, which may account for that discrepancy. For CeMnAsO, the authors of Ref. \cite{Zhang-2015} note that large values of $\beta$ can arise from strongly anisotropic exchange interactions, though it is not clear why CeMnAsO would be more anisotropic than LaMnAsO. Values similar to those found here are reported for BaMnPF, $\beta$\,=\,0.31(1), and BaMnAsF, $\beta$\,=\,0.36(3) \cite{Saparov-2013}.

The SRO decreases upon cooling below this $T_N$, and is nearly absent at 300\,K. Upon heating above $T_N$ the SRO decreases gradually, and vanishes near the temperature at which the magnetic susceptibility anomaly is observed (Fig. \ref{fig:mag}b). This is used to identify the onset of magnetic correlations at $T_{SRO}$\,=\,650(10)\,K.


\begin{figure}
\begin{center}
\includegraphics[width=3.00in]{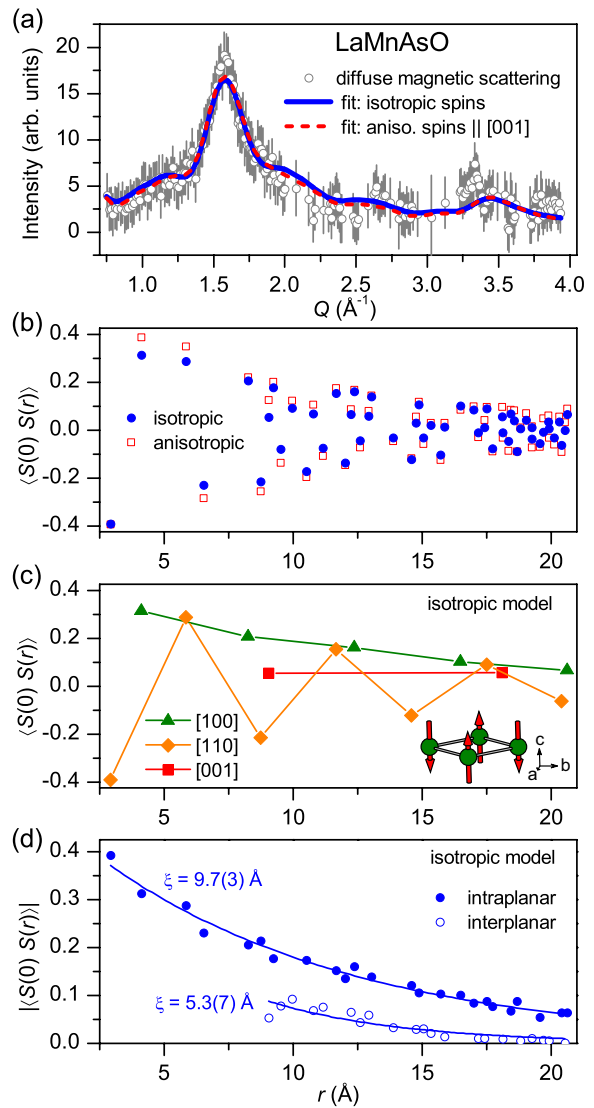}
\caption{\label{fig:diffuse}
(a) Measured diffuse magnetic scattering (circles) and calculated scattering (lines) from the spin configurations determined by reverse Monte Carlo using the program SPINVERT \cite{SPINVERT}. (b) The radial dependence of the spin-pair correlation function corresponding to the fits shown in (a). (c) Data from (b) grouped into sets corresponding spins separated along to the indicated crystallographic directions. (d) Absolute value of the data in (b) for pairs of spins within a common layer (closed symbols) and in neighboring layers (open symbols). The vertical error bars on the spin-pair correlation values in panels b,c,d are approximately the size of the data markers.
}
\end{center}
\end{figure}

To allow further analysis of the SRO in LaMnAsO, the diffuse scattering intensity was extracted from the neutron powder diffraction pattern collected at 400\,K. This was accomplished by using the background determined at 300\,K together with a fit of the nuclear scattering profile at 400\,K. The resulting difference curve reflects the broad diffuse scattering present at 400\,K, as well as sharp features corresponding to errors in the fitted intensities of the Bragg peaks. The sharp features were excluded and the resulting data is shown in Figure \ref{fig:diffuse}a.

The diffuse scattering data was analyzed using SPINVERT \cite{SPINVERT}. This program uses a reverse Monte Carlo algorithm to fit non-periodic spin configurations to measured neutron powder diffraction data. A supercell of the known nuclear structure is generated and a randomly oriented magnetic moment is assigned to each site occupied by a magnetic ion, in this case Mn. A 10$\times$10$\times$10 supercell was used here. The scattering from the spin configuration is calculated and compared with the experimental data. The orientation of a spin is then changed and a determination is made of whether the change improved the fit or not. If it is improved, the new configuration is accepted and the process repeats until a convergence criterion is met or a maximum number of ``moves'' are performed. This is typically done multiple times with different starting configurations to ensure the robustness of the results. Here five runs were performed with a maximum of 300 moves per spin in each run. For more details see Ref. \citenum{SPINVERT} and references therein.

The diffuse scattering intensity calculated from the resulting spin configurations is shown in Figure \ref{fig:diffuse}a as a solid line for an isotropic model and as a dashed line for an anisotropic model with spins directed along the \textit{c}-axis. The results are nearly identical for the two models, and reproduce the measured intensity well. The spin-pair correlation function $\langle S(0) S(r)\rangle$ is shown for both models in Figure \ref{fig:diffuse}b, where each data point corresponds to a distance between two Mn sites within the LaMnAsO crystal structure. Larger absolute values of this function indicate stronger preference for collinear arrangement of the spins separated by the distance \textit{r}, and the sign indicates whether the spins are parallel (+) or antiparallel (-). Again, there is little difference observed between the results of the isotropic and anisotropic spins models, and further analysis will focus only on the isotropic results.

The spin-pair correlations shown in Figure \ref{fig:diffuse}b are strongly diminished as \textit{r} increases up to 20\,{\AA}, and include both positive and negative values. This is indicative of short-range antiferromagnetic correlations. Based on the interatomic distance, each data point can be identified with a specific pair of Mn atoms within the LaMnAsO structure. For reference, a single Mn square is shown in the inset of Figure \ref{fig:diffuse}c.  The shortest Mn$-$Mn distance is along the edge of the Mn square lattice lying in the \textit{ab}-plane. Thus, since the Mn square lattice is rotated by 45 degrees with respect to the square face of the tetragonal crystal lattice, the nearest Mn$-$Mn distance is along the [110] direction and equal to $a/\sqrt{2}$\,=\,2.92\,{\AA}, and the next nearest Mn$-$Mn distance, which is the diagonal of the Mn square lattice, is in the [100] direction and equal to \textit{a}\,=\,4.12\,{\AA}. By grouping together points in Figure \ref{fig:diffuse}b that are at multiples of these distances, spin-pair correlations along [100] and [110] can be separately isolated. The results are shown in Figure \ref{fig:diffuse}c. It can be seen from this Figure that ferromagnetic alignment of the spins is preferred along the [100] direction, and antiferromagnetic alignment of the spins is preferred along the [110] direction. Distances along the \textit{c} axis, or [001], are also shown. The correlations for these pairs of spins are positive and significantly smaller than the in-plane correlations. The signs of magnetic interactions along these three crystallographic directions are the same as those observed in the LRO state below $T_N$ (Fig. \ref{fig:npd}a). The reduced magnitude of $\langle S(0) S(r)\rangle$ along [001] relative to [100] and [110] shows that the magnetic interactions are relatively weak between the Mn layers.

The correlation length associated with the SRO can be defined by how rapidly the $\langle S(0) S(r)\rangle$ decrease with \textit{r}. For this purpose, the absolute value of the spin-pair correlation function is examined. Figure \ref{fig:diffuse}d displays the data for pairs that lie within a Mn layer (intraplanar, solid symbols) and pairs which lie within two neighboring Mn layers, separated by one unit cell along the \textit{c} axis (interplanar, open symbols). These sets form two well-separated branches on the $|\langle S(0) S(r)\rangle|$ plot. Each is fitted with an exponential decay toward zero $|\langle S(0) S(r)\rangle| = Ae^{-r/\xi}$.  The resulting correlation lengths $\xi$ are shown on the figure. As expected, the correlation length within the layers is larger than that between the layers. The values are consistent with that approximated by the width of the strongest diffuse scattering feature in Figure \ref{fig:npd}d. In the \textit{ab} plane, $\xi$ corresponds to approximately two unit cells. For pairs of spins in neighboring layers, the magnitude of $\xi$ is likely less reliable, since it is smaller than the smallest distance used to determine it. However, it does indicate weaker magnetic correlations along the \textit{c} axis, as may be expected. In the related compound BaMn$_2$As$_2$, which is structurally less anisotropic than LaMnAsO, exchange interactions along the layer stacking direction have been estimated to be about 10 times weaker than in-plane interactions \cite{Johnston-2011}.

This analysis of the diffuse magnetic scattering reveals that the magnetic order in LaMnAsO is short-range and nearly two dimensional between 650 and 360\,K. It is interesting to consider whether similar magnetic correlations exist above the long-range ordering temperature in related materials, including other Mn based 1111 and 122 compounds. Several examples can be found in the literature. Wildman \textit{et al.} noted some diffuse intensity attributed to SRO at 350\,K in PrMnAsO$_{0.95}$F$_{0.05}$ ($T_N$\,=\,340\,K) \cite{Wildman-2015}. Similarly, in CeMnAsO ($T_N$\,=\,347\,K) a broad magnetic peak was observed up to about 420\,K \cite{Zhang-2015}. In addition, neutron diffraction data collected at 600\,K from Ba$_{0.75}$K$_{0.25}$Mn$_2$As$_2$ ($T_N$\,=\,575\,K) \cite{Lamsal-2013} and at 650\,K from BaMn$_2$As$_2$ ($T_N$\,=\,625\,K) \cite{Singh-2009b} show the presence of diffuse scattering. Thus, it appears that SRO above $T_N$ is a common feature of the magnetism in Mn-based 1111 and 122 type compounds, and may have been overlooked in previous studies of the other materials listed in Table \ref{tab:compare} that focused primarily on the long range magnetic order.

\subsection{Neutron scattering: magnetic excitations}

\begin{figure*}
\begin{center}
\includegraphics[width=6.25in]{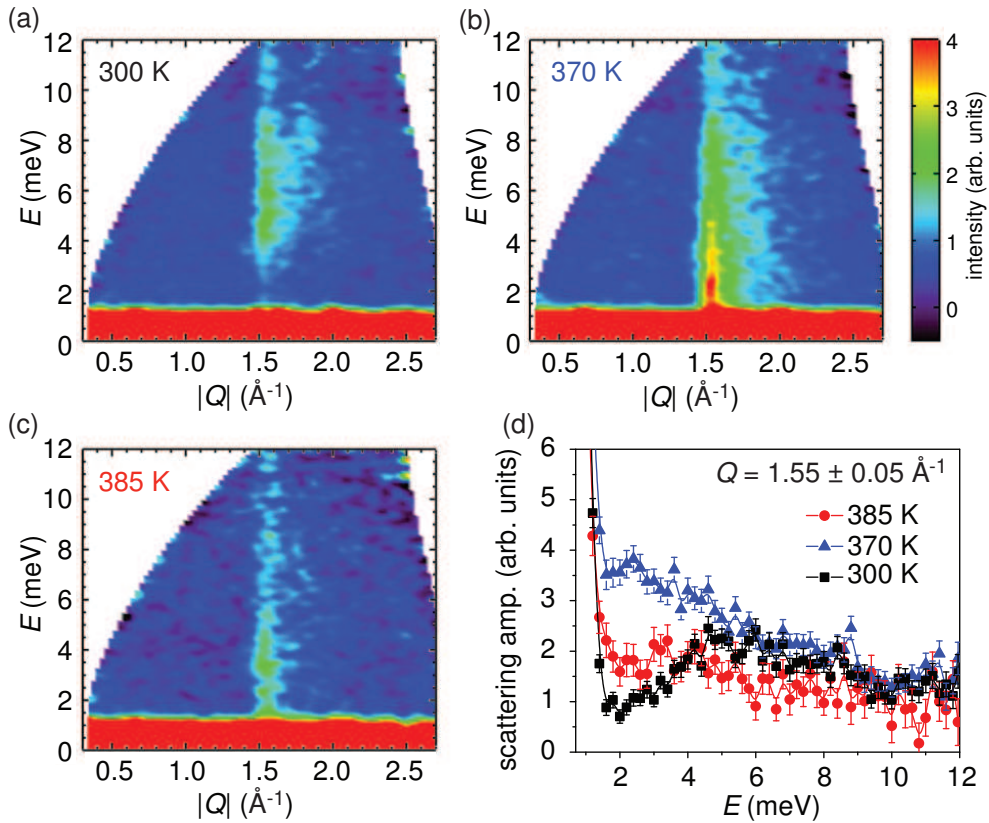}
\caption{\label{fig:INS-maps}
(a-c) Energy versus momentum transfer (\textit{Q}) slices of the polarized inelastic neutron scattering measured in spin-flip configuration at 300\,K, 370\,K and 385\,K. Highly dispersive spin-wave excitation are located at momentum transfer \textit{Q}=1.6{\AA}$^{-1}$ near the 100 and 101 magnetic peak positions. The excitation spectrum at 300\,K displays an anisotropy gap that vanishes with the long-range order. (d) Constant-\textit{Q} cuts near \textit{Q}\,=\,1.55\,{\AA} revealing  a spin-gap of approximately 3.5\,meV in the LRO state, at 300\,K, and the gapless nature of the excitation spectrum in the SRO state at 370\,K.
}
\end{center}
\end{figure*}

Polarized inelastic neutron scattering experiments were performed to examine the magnetic excitation spectrum of LaMnAsO in both the LRO and SRO states. Results of measurements using the spin-flip configuration are shown in Figure \ref{fig:INS-maps}. Measurements at 300\,K, below $T_N$, reveal the presence of well-defined spin-wave excitations located near a momentum transfer $|Q|$\,=\,1.6\,${\AA}^{-1}$. The feature was not observed in the non-spin-flip configuration, demonstrating that it is magnetic in nature. The momentum transfer at which the strong inelastic scattering is centered is close to that where the strongest magnetic Bragg peaks and the maximum in the diffuse magnetic scattering are located (Fig. \ref{fig:diffuse}d). At 300K, a gap is observed in the magnetic excitation spectrum, which arises from magnetic anisotropy in the LRO state. The gap is estimated to be 3.5\,meV from the present data. As noted above, the spectrum is very narrow in \textit{Q}, indicating the spin waves producing this feature are very sharply dispersing. This scattering is quite similar to the polycrstalline averaged spectrum produced from spin-wave calculations on BaMn$_2$As$_2$ \cite{Johnston-2011}. Non-polarized measurements using higher incident energy neutrons showed that the magnetic spectrum in LaMnAsO extends to beyond 30\,meV, and evidence for magnetic excitations extending up to $\sim$100\,meV is seen in BaMn$_2$As$_2$ \cite{Johnston-2011}. The stiff spin wave spectra in these materials indicate strong magnetic exchange interactions and are consistent with the observed high magnetic ordering temperatures.

As the temperature is raised above $T_N$\,=\,360\,K, the inelastic magnetic scattering (Fig. \ref{fig:diffuse}b,c) remains narrow in \textit{Q}; however, the spin-gap vanishes with the long-range order. Just above $T_N$, at 370\,K, very intense quasielastic scattering associated with the short range magnetic order is observed at 1.6\,${\AA}^{-1}$. This quasielastic scattering is suppressed strongly as temperature is increased further, as seen by comparison of the data at 370\,K (Fig. \ref{fig:diffuse}b) and 385\,K (Fig. \ref{fig:diffuse}c). Fixed-\textit{Q} cuts through the spectra are collected in Figure \ref{fig:diffuse}d, which shows the spin-gap at 300\,K, strong, low-energy scattering at 370\,K, and weaker magnetic scattering at 385\,K that still extends up beyond 12\,meV. The scaling of the energy gap with the ordered moment is expected as the the thermal energy overcomes the magnetic anisotropy energy, leading to a gradual decay of the spin correlations and thus to a short range order. However, the strong exchange interaction makes the magnetic excitations persist at temperatures well above the LRO regime.

\section{Summary and Conclusions}

The long range antiferromagnetic order present at room temperature in LaMnAsO vanishes above $T_N$\,=\,360\,K. No corresponding anomaly is detected in the magnetic susceptibility, because short range magnetic order is present above $T_N$. This short range order extends up to $T_{SRO}$\,=\,650\,K. A change is observed in the temperature dependence of the magnetic susceptibility at $T_{SRO}$. Analysis of the diffuse magnetic scattering shows that the magnetic correlations associated with the short range order are anisotropic, with longer correlation length ($\xi\approx$ 10{\AA}) in the \textit{ab}-plane than along the \textit{c}-axis ($\xi\approx$ 5{\AA}). The magnetic exchange is similar to that in the long range ordered state: antiferromagnetic along the edge of the Mn square lattice and ferromagnetic between layers. Comparison of the long range ordered magnetic structures of related compounds shows that the magnetic exchange along the stacking direction is determined by the Mn \textit{d}-orbital crystal field splitting, with a strong correlation between $Pn-$Mn$-Pn$ bond angle and the layer stacking sequence. This suggests chemical substitutions or applied pressure may be used to switch the interlayer magnetic coupling. A gap of about 3.5\,eV in the magnetic excitation spectrum of LaMnAsO is observed at room temperature (below $T_N$), arising from magnetic anisotropy. Steeply dispersing spin waves extend beyond 30\,meV and indicate strong magnetic exchange interactions consistent with high magnetic ordering temperatures. Very strong quasielastic scattering associated with the short range magnetic order is seen just above $T_N$. Comparison with neutron powder diffraction data from other \textit{Ln}MnAsO and BaMn$_2$As$_2$ based materials shows that short ranged order above $T_N$ may be a common fundamental behavior of these quasi-two dimensional magnetic materials.

\section{Acknowledgements}
This work was supported by the U. S. Department of Energy, Office of Science, Basic Energy Sciences, Materials Sciences and Engineering Division (M.A.M.) and Scientific User Facilities Division (V.O.G). The authors are grateful for helpful discussions with S. Calder, B.C. Sales, A.F. May, and J.-Q. Yan.

%\bibliography{LaMnAsO}% Produces the bibliography via BibTeX.

%

\end{document}